\PassOptionsToPackage{unicode}{hyperref}
\PassOptionsToPackage{hyphens}{url}
\PassOptionsToPackage{dvipsnames,svgnames,x11names}{xcolor}
\documentclass[
]{article}
\usepackage{amsmath,amssymb}
\usepackage{lmodern}
\usepackage{iftex}
\ifPDFTeX
  \usepackage[T1]{fontenc}
  \usepackage[utf8]{inputenc}
  \usepackage{textcomp}
\else % if luatex or xetex
  \usepackage{unicode-math}
  \defaultfontfeatures{Scale=MatchLowercase}
  \defaultfontfeatures[\rmfamily]{Ligatures=TeX,Scale=1}
\fi
\IfFileExists{upquote.sty}{\usepackage{upquote}}{}
\IfFileExists{microtype.sty}{% use microtype if available
  \usepackage[]{microtype}
  \UseMicrotypeSet[protrusion]{basicmath} % disable protrusion for tt fonts
}{}
\makeatletter
\@ifundefined{KOMAClassName}{% if non-KOMA class
  \IfFileExists{parskip.sty}{%
    \usepackage{parskip}
  }{% else
    \setlength{\parindent}{0pt}
    \setlength{\parskip}{6pt plus 2pt minus 1pt}}
}{% if KOMA class
  \KOMAoptions{parskip=half}}
\makeatother
\usepackage{xcolor}
\newlength{\cslhangindent}
\newlength{\csllabelwidth}
\newlength{\cslentryspacingunit} % times entry-spacing
\newenvironment{CSLReferences}[2] % #1 hanging-ident, #2 entry spacing
 {% don't indent paragraphs
  \setlength{\parindent}{0pt}
  \ifodd #1
  \let\oldpar\par
  \def\par{\hangindent=\cslhangindent\oldpar}
  \fi
  \setlength{\parskip}{#2\cslentryspacingunit}
 }%
 {}
\usepackage{calc}

\ifLuaTeX
\usepackage[bidi=basic]{babel}
\else
\usepackage[bidi=default]{babel}
\fi
\babelprovide[main,import]{american}
\def\languageshorthands#1{}
\ifLuaTeX
  \usepackage{selnolig}  % disable illegal ligatures
\fi
\IfFileExists{bookmark.sty}{\usepackage{bookmark}}{\usepackage{hyperref}}
\IfFileExists{xurl.sty}{\usepackage{xurl}}{} % add URL line breaks if available
\hypersetup{
  pdftitle={libcdict: fast dictionaries in C},
  pdfauthor={Robert G. Izzard, David D. Hendriks, Daniel P. Nemergut},
  pdflang={en-US},
  colorlinks=true,
  linkcolor={Maroon},
  filecolor={Maroon},
  citecolor={Blue},
  urlcolor={Blue},
  pdfcreator={LaTeX via pandoc}}

\title{libcdict: fast dictionaries in C}

\usepackage[affil-it]{authblk}
\usepackage{orcidlink}
\author[1%
  \ensuremath\mathparagraph]{Robert G. Izzard%
    \,\orcidlink{0000-0003-0378-4843}\,%
    }
\author[1%
  ]{David D. Hendriks%
    \,\orcidlink{0000-0002-8717-6046}\,%
    }
\author[1%
  ]{Daniel P. Nemergut%
    \,\orcidlink{0009-0001-5004-7515}\,%
    }

\affil[1]{Department of Physics, School of Mathematics and Physics,
University of Surrey, Guildford, GU2 7XH, Surrey, UK}
\affil[$\mathparagraph$]{Corresponding author}
\date{06 December 2023}

\begin{document}
\maketitle

\hypertarget{summary}{%
\section{Summary}\label{summary}}

A common requirement in science is to store and share large sets of
simulation data in an efficient, nested, flexible and human-readable
way. Such datasets contain number counts and distributions,
i.e.~histograms and maps, of arbitrary dimension and variable type,
e.g.~floating-point number, integer or character string. Modern
high-level programming languages like Perl and Python have associated
arrays, knowns as dictionaries or hashes, respectively, to fulfil this
storage need. Low-level languages used more commonly for fast
computational simulations, such as C and Fortran, lack this
functionality. We present a \texttt{libcdict}, a C dictionary library,
to solve this problem. \texttt{Libcdict} provides C and Fortran
application programming interfaces (APIs) to native dictionaries, called
\texttt{cdict}s, and functions for \texttt{cdict} to load and save these
as JSON and hence for easy interpretation in other software and
languages like Perl, Python and R.

\hypertarget{statement-of-need}{%
\section{Statement of need}\label{statement-of-need}}

Users of high-level languages such as Perl or Python have access to
associated-array data structures through dictionaries and hashes,
respectively. These allow arbitrary data types to be stored in
array-like structures. These are in turn accessed through key-value
pairs which allow the value to be a further, nested associated array,
allowing arbitrary nesting of data. Compiled low-level languages, like C
and Fortran, are more suited to high-speed and repeated calculations
typical in science. These languages lack native associated-array
functionality. While there are pure hash-table solutions out there, such
as \texttt{glib} (\protect\hyperlink{ref-glib_manual_hashtable}{Glib,
2022}) and \texttt{uthash}
(\protect\hyperlink{ref-uthash_github}{Hansen, 2022}), these do not
combine a simple API for setting and adding to nested structures, a
small library footprint, fast input and output, and standardised JSON
output to easily interface with other languages and tools.
\texttt{libcdict} provides an API for such functionality which allows
\texttt{cdict}s to be nested in \texttt{cdict}s, hence
arbitrarily-nested dictionaries of variables in C just as in Perl or
Python.

\texttt{libcdict} is written in C and provides an API through a set of C
macros. Nested \texttt{cdict} structures have values in them set with a
single line of code. \texttt{libcdict} has been used for the last year
in the \texttt{binary\_c} single- and binary-star population
nucleosynthesis framework (\protect\hyperlink{ref-izzard:2004}{Izzard et
al., 2004}, \protect\hyperlink{ref-izzard:2006}{2006},
\protect\hyperlink{ref-izzard:2009}{2009},
\protect\hyperlink{ref-izzard:2018}{2018}). Recent works
(\protect\hyperlink{ref-2023MNRAS.524.4315H}{Hendriks \& Izzard, 2023b};
\protect\hyperlink{ref-izzard:2023}{Izzard \& Jermyn, 2023};
\protect\hyperlink{ref-2023MNRAS.524.3978M}{Mirouh et al., 2023};
\protect\hyperlink{ref-2023MNRAS.tmp.3267Y}{Yates et al., 2023}) compute
the evolution of millions of single- and binary-stellar systems in only
a few hours using its \texttt{binary\_c-python} Python frontend
(\protect\hyperlink{ref-Hendriks:2023}{Hendriks \& Izzard, 2023a}). We
provide \texttt{libcdict} as open-source code on Gitlab subject to the
GPL3. \texttt{libcdict} also has a comprehensive test suite run through
its configuration program \texttt{cdict-config}.

\hypertarget{using-libcdict}{%
\section{Using libcdict}\label{using-libcdict}}

\texttt{libcdict} is flexible but pragmatic. Keys to \texttt{cdict}s can
be any C scalar or pointer. Values can be scalars, pointers, arrays or
other \texttt{cdict}s, but arrays must be of a single C type. Values can
store metadata of arbitrary type. Pointer values are optionally garbage
collected when a \texttt{cdict} is freed. A set of API macros provides
simple nesting facilities so that placing a value in a nested location
given a list of keys is a simple task for the C programmer. Issues such
as C variable typing are automatically handled for the user.

Variables are internally hashed using \texttt{uthash}
(\protect\hyperlink{ref-uthash_github}{Hansen, 2022}). \texttt{libcdict}
provides a custom JSON output function and inputs JSON using
\texttt{jsmn} (\protect\hyperlink{ref-jsmn_github}{Zaitsev, 2022}).
Floating-point input and output uses \texttt{fast\ double\ parser}
(\protect\hyperlink{ref-lemire:2021}{Lemire, 2021},
\protect\hyperlink{ref-fastdoubleparser_github}{2022}) and \texttt{Ryū}
(\protect\hyperlink{ref-adams:2018}{Adams, 2018},
\protect\hyperlink{ref-adams:2019}{2019},
\protect\hyperlink{ref-ryugithub}{2022}), respectively, both of which
are considerably faster than equivalent C library functions.
\texttt{libdict} allows customizable floating-point accuracy for output
and when comparing floating-point numbers, e.g.~during sorting of key or
variable lists.

Installation uses \texttt{meson}
(\protect\hyperlink{ref-meson_github}{Pakkanen, 2022}) and
\texttt{ninja} (\protect\hyperlink{ref-ninja_github}{Martin, 2022}).
\texttt{libcdict} has been tested with the GCC (10.3.0) and Clang
(12.0.0) compilers.

\hypertarget{libcdict-in-stellar-population-statistics-calculations}{%
\section{\texorpdfstring{\texttt{libcdict} in stellar-population
statistics
calculations}{libcdict in stellar-population statistics calculations}}\label{libcdict-in-stellar-population-statistics-calculations}}

\texttt{libcdict} was developed to solve the problem of storing
statistics in stellar-population calculations in \texttt{binary\_c}.
When evolving a population of millions, sometimes billions, of stars,
each for thousands of time steps, enormous amounts of data are computed.
It is impractical to output these data every time step as these are
typically \(\sim 10^{6} \times 10^{4} = 10^{10}\) lines, each of which
can easily be \(\sim 1\,\mathrm{KB}\) long. The data from each star
could be sent to a Perl or Python front-end which merges them into a
dictionary of population statistics. This communication between
programming languages involves significant overhead which compares
similarly to the runtime of the stellar code itself thus greatly
increases runtime and cost.

To overcome this problem, \texttt{binary\_c} internally generates an
associative-array \texttt{cdict} in native C. This \texttt{cdict}, and
the stellar statistics it contains, is filled inside the
\texttt{binary\_c} simulation as each star is simulated. Generation of
the stellar-population data in the \texttt{cdict} is efficient because
it is only in C and communication with the frontend (Python) code is
kept to a minimum. The \texttt{cdict}'s dataset is output
\emph{only once}, as human-readable JSON easily understood by Perl or
Python, at the end of the simulation. Large simulations are often split
across clusters of machines using \texttt{binary\_c-python}. The data
from each run are stored as JSON chunks then merged in Python when the
final run completes. The overhead involved in this joining is small
compared to the effort of simulating the stars: the goal of
\texttt{libcdict} has thus been achieved.

We provide an interactive example made with \texttt{binary\_c} and
\texttt{binary\_c-python} using \texttt{libcdict} in its
\texttt{examples} directory
(\protect\hyperlink{ref-cdictexamples}{Izzard, 2022}). The
\texttt{libcdict} JSON output of a Hertzsprung-Russell diagram, the most
important diagnostic plot in stellar astrophysics, is plotted using
\texttt{Bokeh} (\protect\hyperlink{ref-bokeh_homepage}{Bokeh Development
Team, 2014}; \protect\hyperlink{ref-bokeh_github}{Bokeh~GitHub, 2022})
to provide immediate access to nested data sets.

\hypertarget{acknowledgements}{%
\section{Acknowledgements}\label{acknowledgements}}

RGI acknowledges funding by STFC grants ST/L003910/1, ST/L003910/2 and
ST/R000603/1. DDH acknowledges funding by UKRI/UoS grant H120341A. We
thank the authors of software used by \texttt{libcdict}, especially Troy
Hanson and Arthur O'Dwyer for \texttt{uthash}, Ulf Adams and the
\texttt{Ryū} team, Serge Zaitsev and the \texttt{jsmu} team, and Daniel
Lemire for \texttt{fast\ double\ parser}.

\hypertarget{references}{%
\section*{References}\label{references}}
\addcontentsline{toc}{section}{References}

\hypertarget{refs}{}
\begin{CSLReferences}{1}{0}
\leavevmode\vadjust pre{\hypertarget{ref-adams:2018}{}}%
Adams, U. (2018). Ryū: Fast float-to-string conversion. \emph{SIGPLAN
Not.}, \emph{53}(4), 270--282.
\url{https://doi.org/10.1145/3296979.3192369}

\leavevmode\vadjust pre{\hypertarget{ref-adams:2019}{}}%
Adams, U. (2019). Ryū revisited: Printf floating point conversion. In
\emph{Proceedings of the ACM on Programming Languages} (OOPSLA; Vol. 3,
pp. 1--23). Association for Computing Machinery (ACM).
\url{https://doi.org/10.1145/3360595}

\leavevmode\vadjust pre{\hypertarget{ref-ryugithub}{}}%
Adams, U. (2022). {Ryū} \& {Ryū} {Printf}. In \emph{GitHub repository}.
\url{https://github.com/ulfjack/ryu}

\leavevmode\vadjust pre{\hypertarget{ref-bokeh_homepage}{}}%
Bokeh Development Team. (2014). \emph{Bokeh: Python library for
interactive visualization}. \url{http://www.bokeh.pydata.org}

\leavevmode\vadjust pre{\hypertarget{ref-bokeh_github}{}}%
Bokeh~GitHub. (2022). Bokeh. In \emph{GitHub repository}. GitHub.
\url{https://github.com/bokeh/}

\leavevmode\vadjust pre{\hypertarget{ref-glib_manual_hashtable}{}}%
Glib. (2022). \textsc{glib} hash tables. In \emph{\textsc{glib} official
documentation}. Gitlab. \url{https://docs.gtk.org/glib/index.html}

\leavevmode\vadjust pre{\hypertarget{ref-uthash_github}{}}%
Hansen, T. D. (2022). {uthash}: A hash table for {C} structures. In
\emph{GitHub repository}. \url{https://troydhanson.github.io/uthash/}

\leavevmode\vadjust pre{\hypertarget{ref-Hendriks:2023}{}}%
Hendriks, D. D., \& Izzard, R. G. (2023a). Binary\_c-python: A
python-based stellar population synthesis tool and interface to
binary\_c. \emph{Journal of Open Source Software}, \emph{8}(85), 4642.
\url{https://doi.org/10.21105/joss.04642}

\leavevmode\vadjust pre{\hypertarget{ref-2023MNRAS.524.4315H}{}}%
Hendriks, D. D., \& Izzard, R. G. (2023b). {Mass-stream trajectories
with non-synchronously rotating donors}. \emph{524}(3), 4315--4332.
\url{https://doi.org/10.1093/mnras/stad2077}

\leavevmode\vadjust pre{\hypertarget{ref-cdictexamples}{}}%
Izzard, R. G. (2022). \textsc{libcdict} examples. In \emph{Gitlab.com
repository}. Gitlab.com. \url{https://doi.org/10.5281/zenodo.10276619}

\leavevmode\vadjust pre{\hypertarget{ref-izzard:2006}{}}%
Izzard, R. G., Dray, L. M., Karakas, A. I., Lugaro, M., \& Tout, C. A.
(2006). {Population nucleosynthesis in single and binary stars. I.
Model}. \emph{Astronomy and Astrophysics}, \emph{460}, 565--572.
\url{https://doi.org/10.1051/0004-6361:20066129}

\leavevmode\vadjust pre{\hypertarget{ref-izzard:2009}{}}%
Izzard, R. G., Glebbeek, E., Stancliffe, R. J., \& Pols, O. R. (2009).
{Population synthesis of binary carbon-enhanced metal-poor stars}.
\emph{Astronomy and Astrophysics}, \emph{508}, 1359--1374.
\url{https://doi.org/10.1051/0004-6361/200912827}

\leavevmode\vadjust pre{\hypertarget{ref-izzard:2023}{}}%
Izzard, R. G., \& Jermyn, A. S. (2023). {Circumbinary discs for stellar
population models}. \emph{521}(1), 35--50.
\url{https://doi.org/10.1093/mnras/stac2899}

\leavevmode\vadjust pre{\hypertarget{ref-izzard:2018}{}}%
Izzard, R. G., Preece, H., Jofre, P., Halabi, G. M., Masseron, T., \&
Tout, C. A. (2018). {Binary stars in the Galactic thick disc}.
\emph{Monthly Notices of the Royal Astronomical Society}, \emph{473},
2984--2999. \url{https://doi.org/10.1093/mnras/stx2355}

\leavevmode\vadjust pre{\hypertarget{ref-izzard:2004}{}}%
Izzard, R. G., Tout, C. A., Karakas, A. I., \& Pols, O. R. (2004). {A
new synthetic model for asymptotic giant branch stars}. \emph{Monthly
Notices of the Royal Astronomical Society}, \emph{350}, 407--426.
\url{https://doi.org/10.1111/j.1365-2966.2004.07446.x}

\leavevmode\vadjust pre{\hypertarget{ref-lemire:2021}{}}%
Lemire, D. (2021). {Number Parsing at a Gigabyte per Second}.
\emph{arXiv e-Prints}, arXiv:2101.11408.
\url{https://doi.org/10.1002/spe.2984}

\leavevmode\vadjust pre{\hypertarget{ref-fastdoubleparser_github}{}}%
Lemire, D. (2022). {fast\_double\_parser}: \(4\times\) faster than
\texttt{strtod}. In \emph{GitHub repository}.
\url{https://github.com/lemire/fast_double_parser}

\leavevmode\vadjust pre{\hypertarget{ref-ninja_github}{}}%
Martin, E. (2022). {Ninja}, a small build system with a focus on speed.
In \emph{GitHub repository}. GitHub.
\url{https://github.com/ninja-build/ninja}

\leavevmode\vadjust pre{\hypertarget{ref-2023MNRAS.524.3978M}{}}%
Mirouh, G. M., Hendriks, D. D., Dykes, S., Moe, M., \& Izzard, R. G.
(2023). {Detailed equilibrium and dynamical tides: impact on
circularization and synchronization in open clusters}. \emph{524}(3),
3978--3999. \url{https://doi.org/10.1093/mnras/stad2048}

\leavevmode\vadjust pre{\hypertarget{ref-meson_github}{}}%
Pakkanen, J. (2022). The {Meson} build system. In \emph{GitHub
repository}. GitHub. \url{https://github.com/mesonbuild/meson}

\leavevmode\vadjust pre{\hypertarget{ref-2023MNRAS.tmp.3267Y}{}}%
Yates, R. M., Hendriks, D., Vijayan, A. P., Izzard, R. G., Thomas, P.
A., \& Das, P. (2023). {The impact of binary stars on the dust and metal
evolution of galaxies}. \url{https://doi.org/10.1093/mnras/stad3419}

\leavevmode\vadjust pre{\hypertarget{ref-jsmn_github}{}}%
Zaitsev, S. (2022). {jsmn}, a minimalistic {JSON} parser in {C}. In
\emph{GitHub repository}. \url{https://github.com/zserge/jsmn}

\end{CSLReferences}

\end{document}